\documentclass[showpacs,preprintnumbers,aps]{revtex4}

\usepackage{latexsym} 
\usepackage{amssymb}  
\usepackage{amsfonts} 
\usepackage{amsbsy}   
\usepackage{amsmath} 
\usepackage{graphicx}       
\usepackage{multirow}  
\usepackage[dvips]{color}
\usepackage{bm}   

\newcommand\eqn[1]{(\ref{#1})}      

\newcommand{\bea}{\begin{eqnarray}}
\newcommand{\eea}{\end{eqnarray}}

\newcommand{\Det}{\mbox{Det}}

\newcommand{\nn}{\nonumber \\}

\makeatletter 

\def\appendix{\par                              
    \setcounter{section}{0}                     
    \setcounter{subsection}{0}
    \renewcommand{\theequation}{\Alph{section}.\arabic{equation}}
    \renewcommand{\thesection}{Appendix \Alph{section}
                \setcounter{equation}{0}  } 
    \renewcommand{\thesubsection}{\Alph{section}.\arabic{subsection}}
}

\def\applabel#1{\@bsphack
  \protected@write\@auxout{}%
         {\string\newlabel{#1}{{\Alph{section}}{\thepage}}}%
  \@esphack}

\def\section{
\setcounter{equation}{0}        
\@startsection {section}{1}{\z@}{-3.5ex plus -1ex minus
 -.2ex}{2.3ex plus .2ex}{\large\bf}}
\renewcommand{\theequation}{\arabic{section}.\arabic{equation}}

\def\subsection{\@startsection{subsection}{2}{\z@}{-3.25ex plus -1ex minus
 -.2ex}{1.5ex plus .2ex}{\normalsize\bf}}

\def\subsubsection{\@startsection{subsubsection}{3}{\z@}{-3.25ex plus
 -1ex minus -.2ex}{1.5ex plus .2ex}{\normalsize}}

\makeatother   

\begin{document}


\title{\bf Fluctuations about Cosmological Instantons}

\author{Gerald V.\ Dunne}
\author{Qing-hai Wang}
\affiliation{Department of Physics, University of Connecticut,
Storrs, CT 06269-3046, USA}


\begin{titlepage}

\begin{abstract}
We study the semiclassical fluctuation problem around bounce
solutions for a self-interacting scalar field in curved space. As
in flat space, the fluctuation problem separates into partial
waves labeled by an integer $l$, and we determine the  large $l$
behavior of the fluctuation determinants, a quantity needed to
define a finite fluctuation prefactor. We also show that while the
Coleman-De Luccia bounce solution has a single negative mode in
the $l=0$ sector, the oscillating bounce solutions also have
negative modes in partial waves higher than the s-wave, further
evidence that they are not directly related to quantum tunneling.
\end{abstract}


\end{titlepage}

\maketitle

\section{Introduction}
\label{sec:intro}

The problem of false vacuum decay in the presence of gravity
\cite{cdl} provides an important window into the behavior of
interacting quantum fields in curved space-time, and is also
important for our understanding of string theory  and quantum
gravity \cite{banks,kachru,bousso1}, and inflationary models of
cosmology \cite{linde}. Since the pioneering work of Coleman and
De Luccia \cite{cdl}, much has been learned regarding the
existence and properties of bounce solutions for interacting
scalar fields coupled to gravity
\cite{hm,jensen,mottola,parke,kimyeong,samuel,banks,demetrian,hw,bousso,banksjohnson},
and  consequently the exponential factor in the false vacuum decay
rate. On the other hand, relatively little is known about the
prefactor in the decay rate. This is in distinct contrast to the
flat space case where the entire computation is well understood
physically and mathematically; analytically in the thin-wall limit
\cite{langer,voloshin,stone,coleman1,voloshin2}, and numerically in general
\cite{konoplich,strumia,baacke,dunnemin}. Here we begin to address
this prefactor question with coupling to gravity by studying the
problem of quantum fluctuations about the  bounce solutions.  A
full solution to this problem is not possible at present for the
simple reason that computing the renormalized fluctuation
prefactor would require an understanding of the renormalization of
quantum gravity. However, we argue that certain interesting things
can be learned, in particular in the limit where the gravitational
background is fixed to be de Sitter space.

In the flat space false vacuum decay problem, the fluctuation
operator separates into partial waves labeled by an integer $l$,
and there are three important types of modes. In the $l=0$ sector
there is a single negative mode and this is responsible for the
decaying nature of the problem
\cite{langer,voloshin,stone,coleman1}. In the $l=1$ sector there
are four zero modes corresponding to translational invariance in
four-dimensional Euclidean space, and these zero modes lead to
collective coordinate contributions to the overall fluctuation
determinant. For $l\geq 2$ the eigenvalues are all positive, and
since for each $l$ the fluctuation operator is a one-dimensional
radial operator, one can compute the determinant straightforwardly
using the Gel'fand-Yaglom method (described below in Section
\ref{largel-sec}). Formally, the determinant of the full
fluctuation operator is a product of the determinants for all $l$,
including degeneracies, so the large $l$ behavior is crucial for
defining a finite renormalized fluctuation determinant. In the
thin-wall limit, where the energy gap between the true and false
vacua is small, the computation can essentially be done
analytically \cite{langer,voloshin,stone,coleman1}, and one has a
beautiful physical picture of this process as nucleation of
bubbles. Away from the thin-wall limit, the computation can be
done by various approximate or numerical approaches
\cite{konoplich,strumia,baacke,dunnemin}. Our main motivation here
is to investigate how the behavior of the fluctuation operator is
affected by the inclusion of coupling to gravity.

With the inclusion of gravity, Coleman and De Luccia argued
\cite{cdl} that the bounce solutions are still radially symmetric
(although this has not been rigorously proved, as it has been in
flat space \cite{coleman2}). Interestingly, new classes of bounce
solutions arise, with different physical interpretations. The
Coleman-De Luccia (CDL) bounce generalizes the flat space bounce
and is presumed to be associated with quantum tunneling
\cite{cdl}. There also exists the Hawking-Moss (HM) bounce
\cite{hm} which is interpreted physically in terms of a thermal
transition \cite{linde,hw}. More recently it has been shown that
there are also ``oscillating bounce'' solutions in which the
scalar field passes over the barrier more than once
\cite{banksjohnson,bousso,demetrian,hw}. As emphasized in
\cite{hw}, these oscillating bounces interpolate between the CDL
and HM bounces, and reflect the thermal character of quantum field
theory in de Sitter space. Since all these bounces are radial, a
similar separation of the fluctuation problem into ``partial
waves'' is possible, with the physically plausible assumption that
such radial fluctuations  dominate. But even with this separation,
the fluctuation problem is still considerably more subtle with the
inclusion of gravity, as it requires a detailed constraint
analysis to disentangle the physical fluctuation fields. Here we
consider the scalar fluctuations in the formalism developed in
\cite{turok1,turok2,lav1,lav2}. The existence of negative modes in
the $l=0$ sector for these scalar fluctuations has been
investigated in \cite{turok1,turok2,lav1,lav2}. We extend this
fluctuation analysis in several ways by considering the behavior
for higher $l$. We study two main questions: First, we investigate
the large $l$ behavior of the fluctuation determinants within each
partial wave sector. We find an explicit expression for the
leading large $l$ behavior, and a  numerically accurate estimate
for the subleading behavior. Second, we analyze the existence of
negative modes not just in the $l=0$ sector, but also for higher
$l$, and show that the oscillating bounce solutions have negative
modes for higher $l$. This is further evidence for the physical
picture in \cite{hw} that these oscillating bounces are not
directly related to quantum tunneling, but rather reflect the
thermal nature of quantum field theory in de Sitter space. We are
not able to study the $l=1$ sector, as this fluctuation formalism
does not apply here \cite{turok1,turok2}, and so this requires a
separate study.

In Section \ref{bounce-sec} we review the model and the
construction of bounce solutions to the classical equations of
motion. In Section \ref{fluc-sec} we summarize the scalar
fluctuation problem to be studied. Section \ref{largel-sec} is
devoted to the study of the large $l$ behavior of the fluctuation
determinant, in which we review the flat space approach. In
Section \ref{neg-sec} we count the negative modes for various $l$
for fluctuations about bounce solutions. Section
\ref{conclusions-sec} contains our conclusions and an Appendix
gives the relation of our fluctuation operator to other forms
considered in the literature.

\section{Classical bounce solutions}
\label{bounce-sec}

Before discussing quantum fluctuations, we briefly review the
derivation of the bounce solutions themselves. We consider the
four dimensional self-interacting scalar field system with
Euclidean action
\begin{equation}
S_E=\int d^4 x
\sqrt{g}\left[\frac{1}{2}\nabla_\mu\phi\nabla^\mu\phi +
V(\phi)-\frac{1}{2\kappa}R\right]\quad, \label{action}
\end{equation}
where the gravitational coupling is expressed as
$\kappa=8\pi/M_{\rm pl}^2$. In terms of the proper time $\sigma$,
the metric has the form
\begin{equation}
ds^2=d\sigma^2+a^2(\sigma)d\Omega_3^2\quad.
\label{metric}
\end{equation}
The classical Euclidean equations of motion are
\begin{eqnarray}
\ddot{\phi}+3\frac{\dot{a}}{a}\dot{\phi}-V'(\phi)&=&0 \quad,
\label{phieq}\\
\dot{a}^2 - \frac{\kappa a^2}{3}\left[\frac{1}{2}\dot{\phi}^2 -
V(\phi)\right] &=& K\quad, \label{aeq}
\end{eqnarray}
where the overdot denotes $\frac{d}{d\sigma}$, and
$V'(\phi)\equiv\frac{\delta V(\phi)}{\delta \phi}$.  $K=0,\pm1$
corresponding to flat, closed/open universes, respectively. The
boundary conditions for the bounce solutions are
\begin{eqnarray}
\dot{\phi}(0)=0\quad, &\qquad& \dot{\phi}(\sigma_{\rm
max})=0\quad,\nn a(0)=0\quad, &\qquad& a(\sigma_{\rm max})=0\quad,
\label{bcs}
\end{eqnarray}
where $\sigma_{\rm max}$ is defined by the last equation:
$a(\sigma_{\rm max})=0$.  In this paper  we consider  $K=1$, which
leads to the normalization condition
\begin{equation}
\dot{a}(0)=1\quad.
\label{adot}
\end{equation}

\begin{figure}[tb]
\begin{center}
\includegraphics[width=0.7\textwidth]{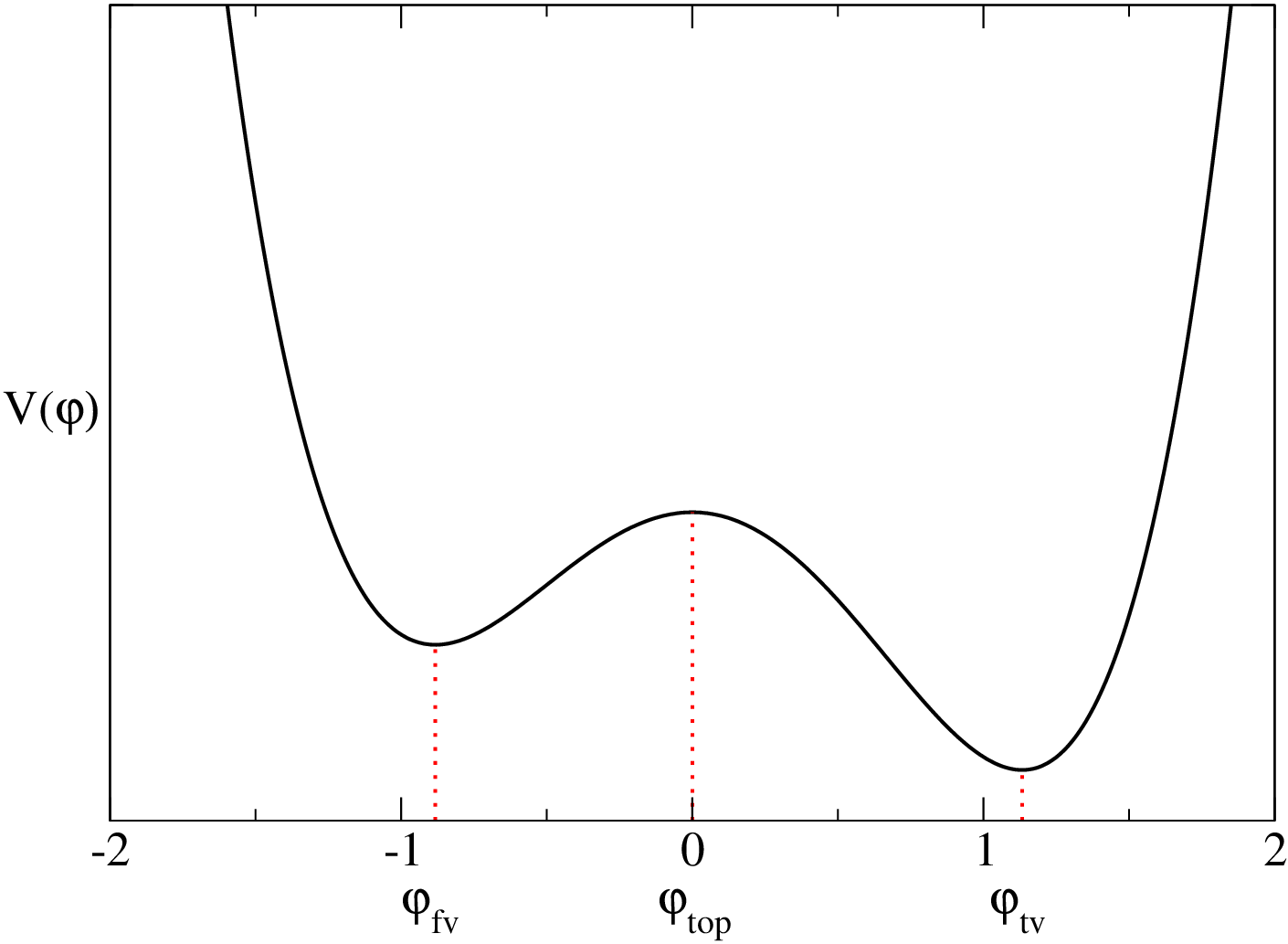}
\end{center}
\caption{The scalar field potential $V(\varphi)$ with the
parameters: $\beta=45$, $b=0.25$, $\epsilon=0.23$.}
\label{fig:Vphi}
\end{figure}

We choose the standard quartic scalar field potential
\cite{hw,banksjohnson}:
\begin{eqnarray}
V(\phi)&=&V_{\rm top}+\beta H_{\rm
top}^2v^2\left[-\frac{1}{2}\left(\frac{\phi}{v}\right)^2
-\frac{b}{3}\left(\frac{\phi}{v}\right)^3
+\frac{1}{4}\left(\frac{\phi}{v}\right)^4\right]\nn&\equiv & \beta
H_{\rm top}^2v^2\left[\frac{1}{\beta
\epsilon^2}+f\left(\frac{\phi}{v}\right)\right]\quad,
\label{potential}
\end{eqnarray}
which is sketched in  Figure \ref{fig:Vphi}.  The function $f$ is
a function of the dimensionless field
$\varphi\equiv\frac{\phi}{v}$:
\begin{equation}
f(\varphi)\equiv -\frac{1}{2}\varphi^2 -\frac{b}{3}\varphi^3 +
\frac{1}{4}\varphi^4\quad. \label{f}
\end{equation}
The potential $V(\phi)$ has two local minima, a false vacuum
$\phi_{\rm fv}$, and a true vacuum $\phi_{\rm tv}$, separated by a
local maximum $\phi_{\rm top}$, chosen to be at $\phi_{\rm
top}=0$. A crucial difference between the flat-space case and the
gravitational case is that the overall constant $V_{\rm top}$ in
the potential is now significant, as it plays the role of a
cosmological constant \cite{cdl}. A corresponding mass scale is
defined as
\begin{equation}
H_{\rm top}\equiv \sqrt{\frac{\kappa V_{\rm
top}}{3}}=\sqrt{\frac{8\pi V_{\rm top}}{3 M_{\rm pl}^2}}\quad.
\label{htop}
\end{equation}
The dimensionless parameter $\beta$ in \eqn{potential}
characterizes the ratio of the barrier curvature (at $\phi_{\rm
top}=0$) to $H_{\rm top}^2$
\begin{eqnarray}
\beta\equiv
\frac{|V^{\prime\prime}(0)|}{H_{\rm top}^2}\quad ,
\label{beta}
\end{eqnarray}
and is an important quantity in determining the existence and
form of bounce solutions \cite{cdl,mottola,jensen,demetrian,hw}.
Another useful dimensionless quantity is the ratio of the field
scale $v$ to the Planck mass $M_{\rm pl}$:
\begin{equation}
\epsilon\equiv \sqrt{\frac{8\pi v^2}{3 M_{\rm
pl}^2}}=\sqrt{\frac{\kappa v^2}{3}}\quad, \label{epsilon}
\end{equation}
and we consider here values such that the potential is everywhere
positive.

The three critical points of $V(\phi)$ correspond to three trivial
solutions to the bounce equations (\ref{phieq}) - (\ref{adot}), in
which $\phi$ is constant at one of these critical values $\phi_c$
such that $V'(\phi_c)=0$:
\begin{eqnarray}
\phi(\sigma)&=&\phi_c \nn a(\sigma)&=&\frac{1}{H}\, \sin(H \sigma)
\quad .
\label{simple}
\end{eqnarray}
The three solutions of this form are : (i) the false vacuum
constant solution with $\phi=\phi_{\rm fv}$, and $H_{\rm fv}\equiv
\sqrt{\kappa V(\phi_{\rm fv})/3}$; (ii) the true vacuum constant
solution with $\phi=\phi_{\rm tv}$, and $H_{\rm tv}\equiv
\sqrt{\kappa V(\phi_{\rm tv})/3}$; (iii) the Hawking-Moss
\cite{hm} solution with  $\phi=\phi_{\rm top}=0$, and $H_{\rm
top}$ given by \eqn{htop}.

More interesting are the bounce solutions in which $\phi$ is not
constant. For definiteness, we restrict our attention to bounces
beginning near the true vacuum and ending near the false vacuum:
\begin{eqnarray}
\phi(0)\approx\phi_{\rm tv}\quad, &\qquad& \phi(\sigma_{\rm
max})\approx \phi_{\rm fv}\quad. \label{truefalse}
\end{eqnarray}
Other bounces exist \cite{demetrian,hw} and can be treated with
completely analogous methods. Bounces can be labeled by an integer
$n$ characterizing how many times they cross the barrier. We will
refer to the Coleman-De Luccia bounce \cite{cdl} as a ``single
bounce'' solution, and the $n\geq 2$ bounces will be termed
``oscillating bounces'' \cite{hw}.  In the flat space limit, only
single bounce solutions have finite action. The explicit bounce
solutions can be found numerically by a straightforward shooting
technique, as follows.

\vskip1cm
\begin{figure}[htb]
\begin{center}
\includegraphics[width=0.45\textwidth]{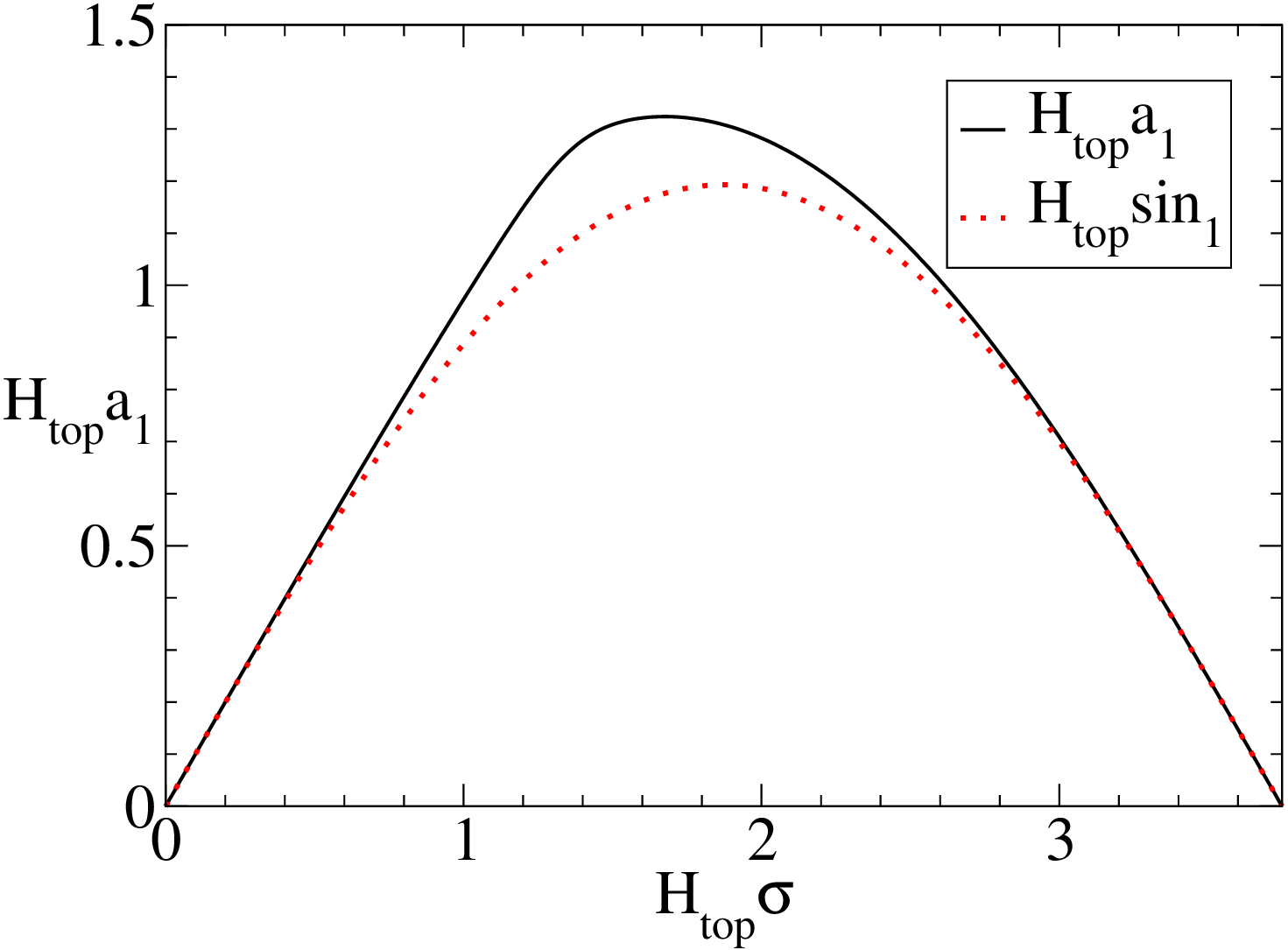}\qquad
\includegraphics[width=0.45\textwidth]{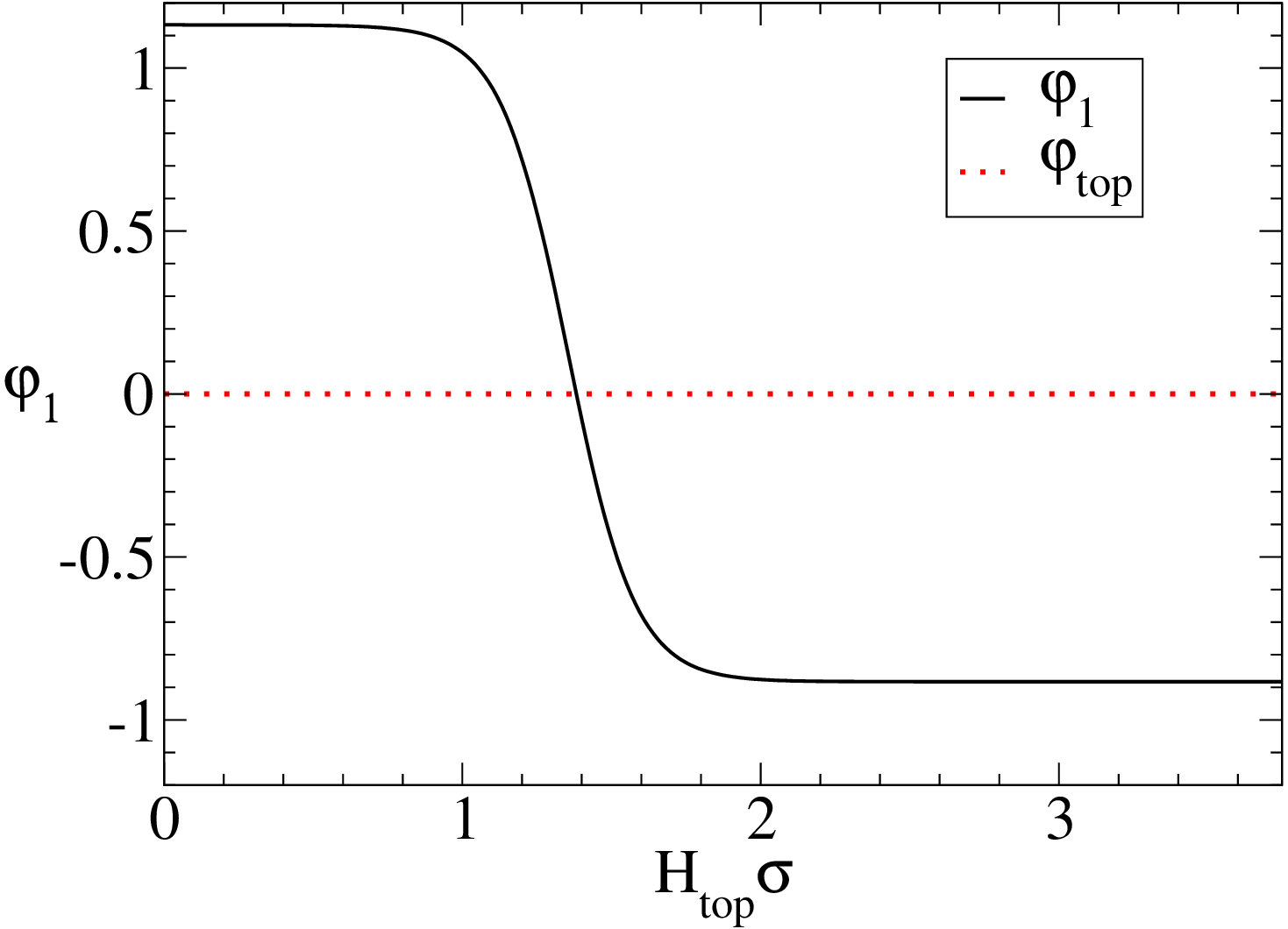}
\end{center}
\caption{Single bounce solution to \eqn{phieq}-\eqn{adot}, for
parameter values: $\beta=45$, $b=0.25$, $\epsilon=0.23$. The
shooting procedure determines the initial scalar field value
$\varphi_1(0)=\varphi_{\rm fv}-0.000\,145\,223\,243\,267\,576\,
9$, and $\sigma_{\rm max}$ is given by: $H_{\rm top}\sigma_{\rm
max}^{\rm 1b}=3.7475$. Here $\sin_1\equiv\frac{1
}{H_1}\sin\left(H_1\sigma\right)$, and $H_1\equiv \pi/\sigma_{\rm
max}^{\rm 1b}$.} \label{fig:phi1}
\end{figure}

\vskip1cm
\begin{figure}[ht]
\begin{center}
\includegraphics[width=0.45\textwidth]{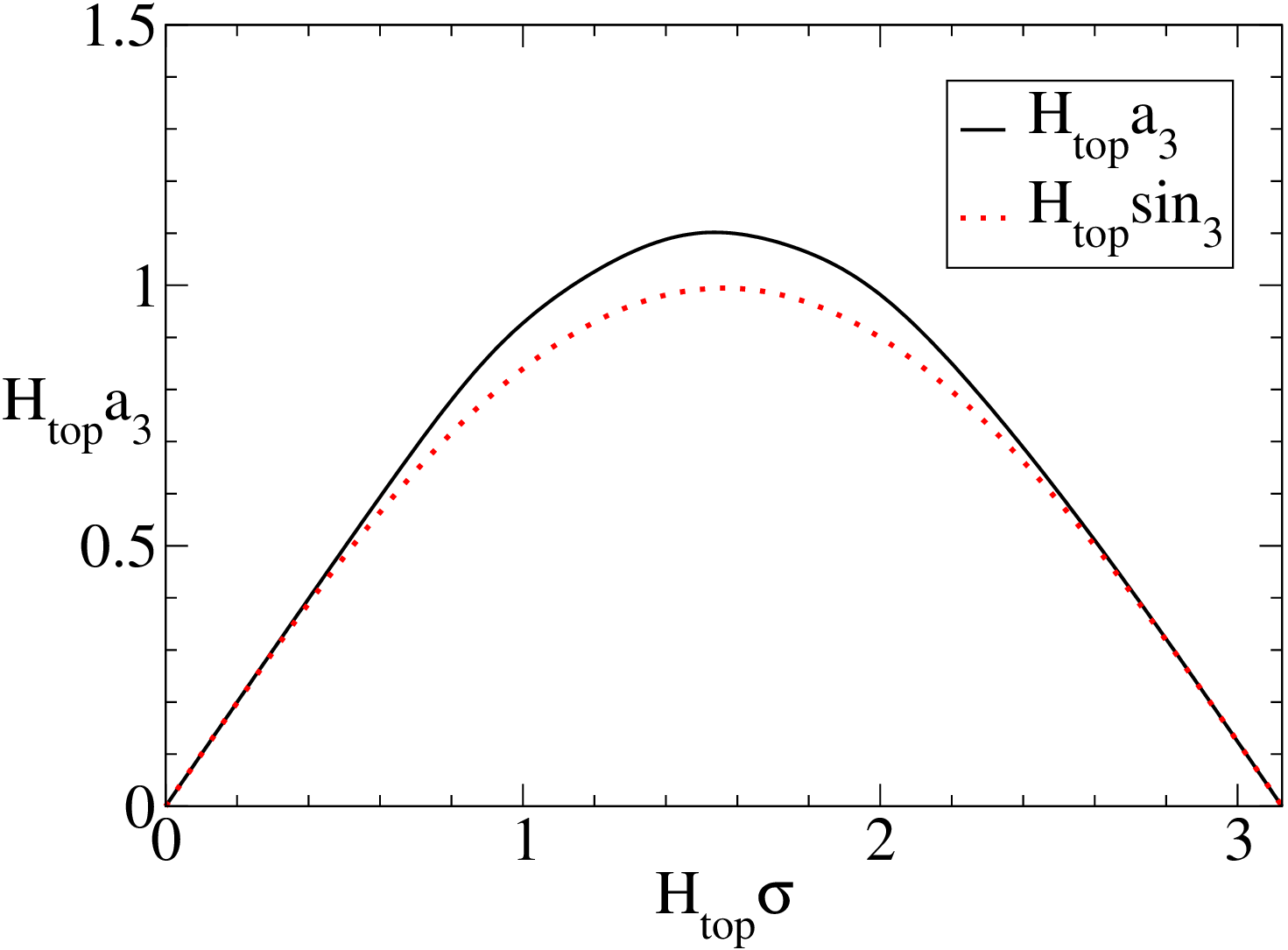}\qquad
\includegraphics[width=0.45\textwidth]{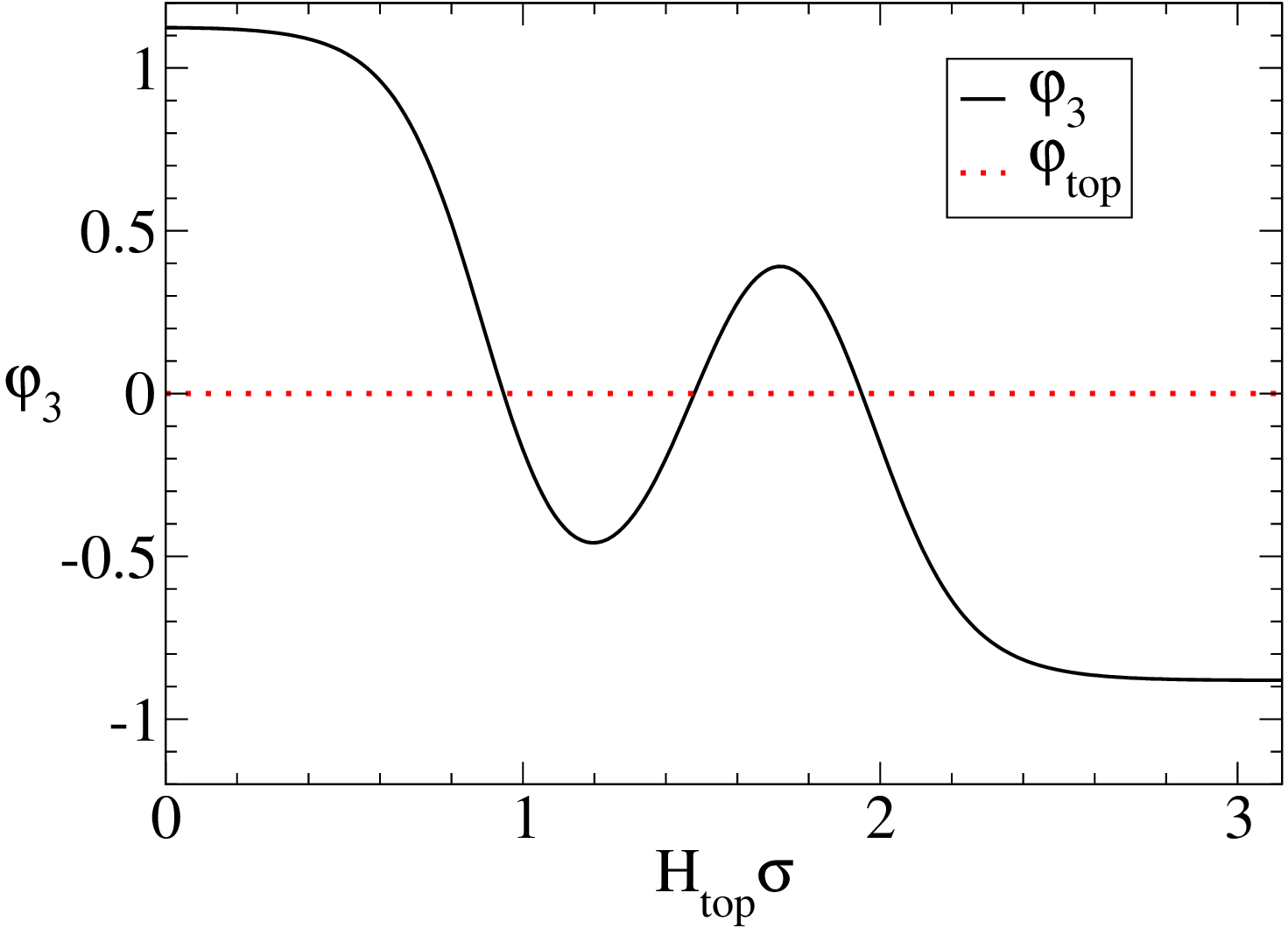}
\end{center}
\caption{Triple bounce solution to \eqn{phieq}-\eqn{adot}, for
parameter values:  $\beta=45$, $b=0.25$, $\epsilon=0.23$. The
shooting procedure determines the initial scalar field value
$\varphi_3(0)=\varphi_{\rm tv}-0.008\,860\,088\,903\,713\,227$,
$H_{\rm top}\sigma_{\rm max}^{\rm 3b}=3.1243$. Here
$\sin_3\equiv\frac{1 }{H_3}\sin\left(H_3\sigma\right)$, and
$\sigma_{\rm max}$ is given by: $H_3\equiv \pi/\sigma_{\rm
max}^{\rm 3b}$.} \vskip1cm \label{fig:phi3}
\end{figure}

\begin{figure}[ht]
\begin{center}
\includegraphics[width=0.45\textwidth]{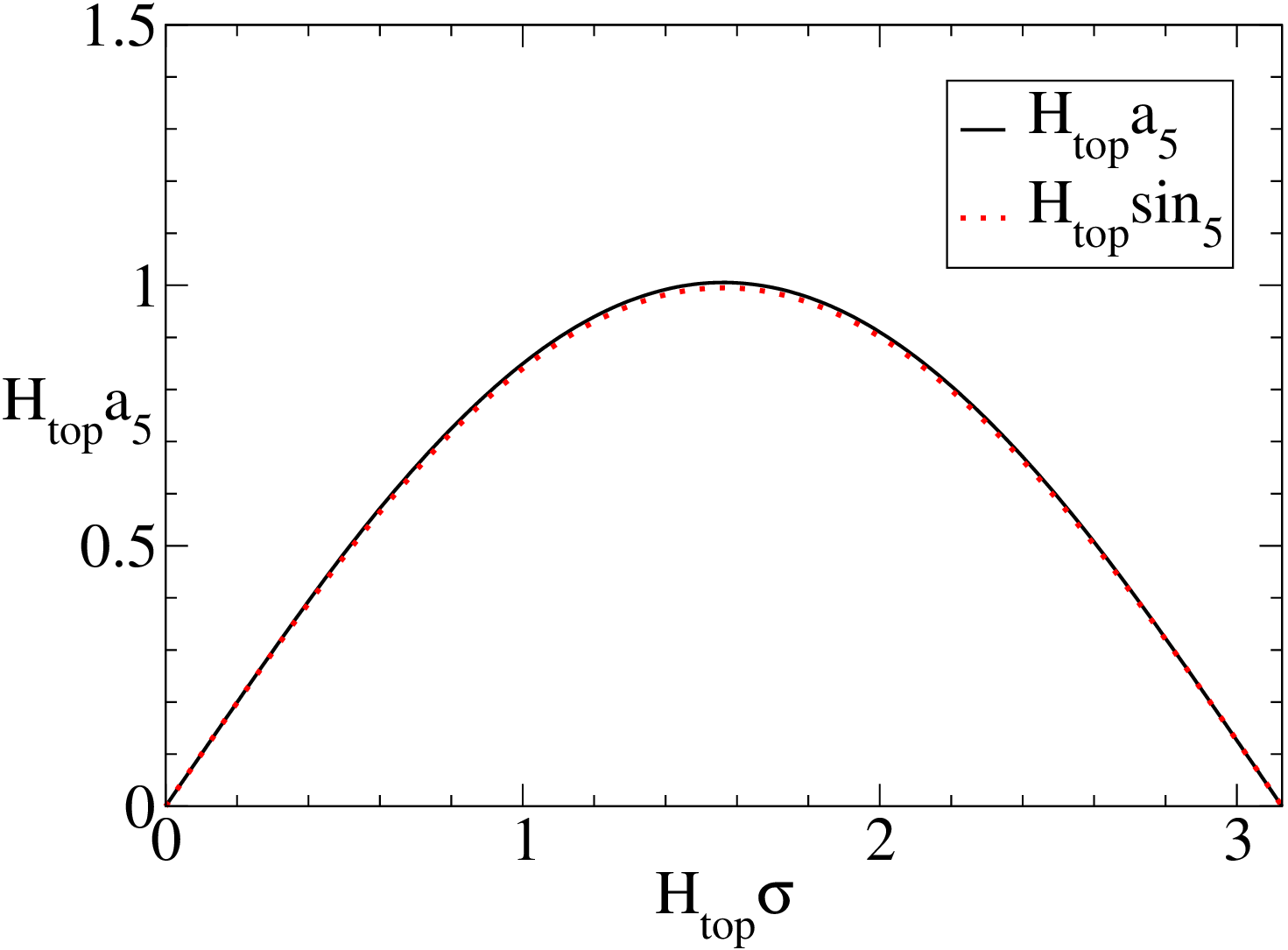}\qquad
\includegraphics[width=0.45\textwidth]{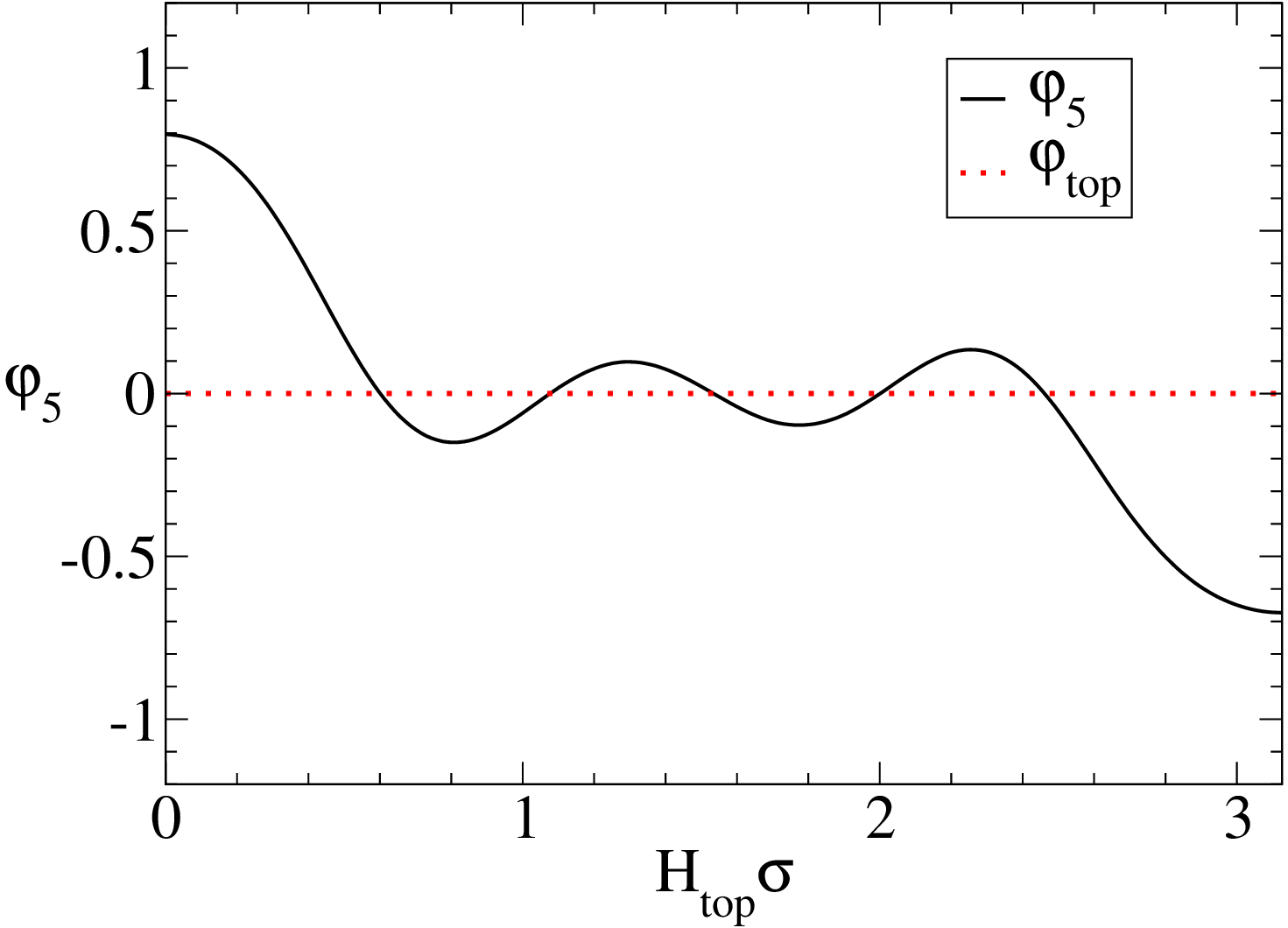}
\end{center}
\caption{Quintuple bounce solution to \eqn{phieq}-\eqn{adot}, for
parameter values:  $\beta=45$, $b=0.25$, $\epsilon=0.23$. The
shooting procedure determines the initial scalar field value
$\varphi_5(0)=\varphi_{\rm tv}-0.337\,102\,748\,195\,264\,1$,
$H_{\rm top}\sigma_{\rm max}^{\rm 5b}=3.1264$. Here
$\sin_5\equiv\frac{1 }{H_5}\sin\left(H_5\sigma\right)$, and
$\sigma_{\rm max}$ is given by: $H_5\equiv \pi/\sigma_{\rm
max}^{\rm 5b}$.} \label{fig:phi5}
\end{figure}

The shooting parameter is the initial value $\phi(0)$ of the
scalar field. This value is chosen near the true vacuum value and
adjusted until the coupled initial value problem \eqn{phieq} -
\eqn{adot} has a solution satisfying both $a(\sigma_{\rm max})=0$
and $\dot{\phi}(\sigma_{\rm max})=0$, for some $\sigma_{\rm max}$.
The value of $\sigma_{\rm max}$ is determined by this shooting
procedure, and so depends on the bounce. Since the metric field
behaves as $a(\sigma)\sim\sigma$ for small $\sigma$, we cannot
directly start integrating \eqn{phieq} at $\sigma=0$. Instead, we
Taylor expand both $a(\sigma)$ and $\phi(\sigma)$ about $0$, and
use these Taylor expansions to begin the integration at a point
very close to $0$. We then do a shooting scan of $\phi(0)$,
adjusting it digit by digit, in a rational form to preserve
precision. This computation is simple to implement in Mathematica.
In a few minutes one can determine $\phi(0)$ to 32 decimal places.
We found that the shooting went faster with the following simple
rescaling of the differential equations, as in
\cite{banksjohnson}: we rescale $\varphi=\phi/v$, and $\sigma$ and
$a(\sigma)$ as: $s=\sqrt{\beta H_{\rm top}^2}\,\sigma$, and $
\rho(s)=\sqrt{\beta H_{\rm top}^2}\,a(\sigma)$, so that the
parameters $v$ and $H_{\rm top}$ scale out of the classical
equations of motion.

Representative examples of bounce solutions with $\phi(0)\approx
\phi_{\rm tv}$ and $\phi(\sigma_{\rm max})\approx\phi_{\rm fv}$
are shown in Figures \ref{fig:phi1},  \ref{fig:phi3} and
\ref{fig:phi5}. These plots are for $\beta=45$. For larger $\beta$
more oscillations are possible in the bounce solutions. Note that
the metric field $a(\sigma)$ deviates significantly from the
sinusoidal form of \eqn{simple} for the single and triple bounce,
but less so for the quintuple bounce. This, together with the fact
that the scalar field $\phi$ is closer to the Hawking-Moss
constant value of $\phi_{\rm top}=0$, is another reflection of the
fact that the highly oscillating bounce solutions tend towards the
Hawking-Moss solution \cite{hw}.

\section{Fluctuation operator}
\label{fluc-sec}

Having reviewed how to find the classical bounce solutions, we now
turn to the problem of fluctuations about these solutions. With
the inclusion of gravity, the fluctuation problem becomes more
subtle, and the fluctuation operator about such cosmological
instantons has been widely studied
\cite{tanaka,turok1,turok2,lav1,lav2}. The variation of the action
under fluctuations of the scalar field and the metric field
requires a nontrivial constraint analysis, with different possible
gauge fixing procedures. For the purpose of this paper, we choose
the gauge fixing scheme described in Section 4.3 of  \cite{lav2},
and Section IV of  \cite{turok2}. In this scheme, the only
physical degree of freedom is the fluctuation of the scalar field,
and the second variation of the action can be expressed as
(compare with Eqn.~(12) in \cite{turok2} and Eqn.~(18) in the
second reference of \cite{lav3}):
\begin{equation}
S^{(2)}[\delta \phi]=\pi^2\int d\sigma\, \delta\phi\left\{
-\frac{d}{d\sigma}\left(\frac{a^3(\sigma)}{{\mathcal Q}(\sigma)}\frac{d}{d\sigma}
+a^3(\sigma)
U[a(\sigma),\phi(\sigma)]\right)\right\}\delta\phi\quad.
\label{second}
\end{equation}
Here
\begin{equation}
{\mathcal Q}(\sigma)\equiv 1+\frac{\kappa
a^2(\sigma)\dot{\phi}^2(\sigma)}{2(-\Delta_3 -3K)}\quad, \label{q}
\end{equation}
and
\begin{eqnarray}
U[a, \phi]&\equiv& \frac{1}{{\mathcal Q}}V''(\phi) +
\frac{-\Delta_3}{{\mathcal Q}a^2}+ \kappa\left\{\frac{2\dot{\phi}^2}{{\mathcal Q}} +
\frac{-a^2\left[V'(\phi)\right]^2 + 5 a\dot{a}\dot{\phi}V'(\phi) -
6\dot{a}^2\dot{\phi}^2}{{\mathcal Q}^2(-\Delta_3-3K)}\right\}\quad.
\label{flucpot}
\end{eqnarray}
Here we have temporarily re-instated the $K$
dependence, although in our numerical studies we return to $K=1$,
and $\Delta_3$ is the Laplacian on $S^3$.

To pass from this secondary action to the Jacobi equation
\cite{morse}, a Sturm-Liouville differential equation whose
eigenvalues will define the determinant of the fluctuation
operator, we need to specify the weight function. The weight
function is determined by defining
\begin{eqnarray}
||\delta \phi ||^2\equiv \int d^4 x \sqrt{g} \left(\delta
\phi\right)^2 =2\pi^2 \int d\sigma\, a^3(\sigma) \left[\delta
\phi(\sigma)\right]^2
\label{weight}
\end{eqnarray}
Then in terms of the perturbation function $\Phi\equiv\delta\phi$,
the fluctuation equation (the Jacobi equation \cite{morse}) is
\begin{equation}
-\frac{1}{a^3}\frac{d}{d\sigma}\left(\frac{a^3}{{\mathcal Q}}\frac{d
\Phi}{d\sigma}\right)+U[a,\phi]\Phi=\lambda \Phi\quad,
\label{sturm}
\end{equation}
which is defined on the interval $\sigma\in [0, \sigma_{\rm
max}]$, with Dirichlet boundary conditions, and where $\lambda$
denotes the eigenvalue. The  ``fluctuation potential'' $U[a,\phi]$
is the function in \eqn{flucpot}. The $S^3$ Laplacian $\Delta_3$
appearing in \eqn{q} and \eqn{flucpot}  can be replaced by its
eigenvalue: $-\Delta_3\to l(l+2)$, so we obtain a fluctuation
equation as an ordinary differential equation for each integer
value of $l$ ($l\neq 1$).

In the flat space limit, $\kappa\to 0$, ${\mathcal Q}\to 1$, $a\to \sigma$,
and $\sigma_{\rm max}\to \infty$; in which case we recover the
familiar flat space fluctuation equation \cite{coleman1}
\begin{eqnarray}
&&-\frac{1}{r^3}\frac{d}{d r}\left(r^3\, \frac{d \Phi}{d
r}\right)+U[\phi(r)]\Phi=\lambda \Phi\quad,\nn
&&U[\phi]=V^{\prime\prime}(\phi)+\frac{l(l+2)}{r^2}\quad,
\label{flatsturm}
\end{eqnarray}
where $\sigma$ becomes identified with the Euclidean length $r$,
which ranges from $0$ to $\infty$. Much is known about solutions
to this flat space fluctuation equation \eqn{flatsturm}. Our goal
now is to study some properties of the more general fluctuation
equation in \eqn{sturm}.

For completeness, we note here that for the purposes of discussing
the existence of negative modes it is possible to make other
choices of the weight function, which yield superficially
different-looking Jacobi operators.  In Appendix A we give the
explicit transformation between our choice \eqn{weight} of weight
function and those made in \cite{turok1,turok2,lav1,lav2,lav3}.

\section{Large $l$ behavior of fluctuation determinants}
\label{largel-sec}

Both $\phi(\sigma)$ and $a(\sigma)$ are functions just of the
proper time $\sigma$, so the fluctuation problem separates into
partial waves, which can be labeled by an integer $l$, just as in
the flat space case. Then formally we can write the log
determinant of the fluctuation operator $\Lambda$ as
\begin{eqnarray}
\ln \left(\frac{\Det\left[ \Lambda \right] }{\Det\left[
\Lambda_{\rm free} \right] }\right)=\sum_{l}^\infty (l+1)^2 \ln
\left(\frac{\Det\left[ \Lambda^{(l)}\right]}{\Det\left[
\Lambda^{(l)}_{\rm free}\right]}\right)\qquad ({\rm
formal~!})\quad,
\label{formal}
\end{eqnarray}
where $\Lambda^{(l)}$ is the differential operator in \eqn{sturm},
for each $l$, and the eigenvalue of $-\Delta_3$ is $l(l+2)$, with
degeneracy $(l+1)^2$. This formal expression \eqn{formal} must be
interpreted with caution, because as in the flat space case, for
low $l$ values there may be negative and zero modes. Nevertheless,
for generic $l$, since each $ \Lambda^{(l)}$ is a one-dimensional
differential operator, the determinant ratio $\left(\Det\left[
\Lambda^{(l)}\right]/{\Det\left[ \Lambda^{(l)}_{\rm
free}\right]}\right)$ is finite, and can be computed efficiently
using the Gel'fand-Yaglom technique
\cite{gy,levit,forman,kirsten,kleinert}. In this approach one
simply numerically integrates both Jacobi equations,
$\Lambda^{(l)}\, \Phi^{(l)}=0$ and $\Lambda^{(l)}_{\rm free}\,
\Phi_{\rm free}^{(l)}=0$, for zero eigenvalue and with suitable
common initial value boundary conditions,
$\Phi^{(l)}(\sigma)\sim\sigma^l$ at $\sigma=0$. Then the ratio of
the determinants is the ratio of these two functions evaluated at
$\sigma_{\rm max}$:
\begin{eqnarray}
\frac{\Det\left[ \Lambda^{(l)}\right]}{\Det\left[
\Lambda^{(l)}_{\rm free}\right]} = \frac{\Phi^{(l)}(\sigma_{\rm
max})}{\Phi^{(l)}_{\rm free}(\sigma_{\rm max})}\quad .
\label{gy}
\end{eqnarray}
This technique provides a simple computational method for
evaluating the finite determinant for each $l$, without ever
having to compute any eigenvalues. However, of course, even though
each term on the RHS of  \eqn{formal} is finite, the sum over $l$
diverges \cite{forman}. This is clearly because we have not
regularized and renormalized the determinant. This divergence is
not a feature of the gravitational coupling --- exactly the same
thing happens in flat space \cite{baacke,dunnemin}, where one can
indeed extract a finite renormalized determinant by subtracting
certain known contributions from $ \ln \left(\Det\left[
\Lambda^{(l)}\right]/{\Det\left[ \Lambda^{(l)}_{\rm
free}\right]}\right)$ for each $l$, rendering the sum finite. The
precise form of the subtractions can be found in various ways,
using Feynman diagram techniques \cite{baacke}, zeta function
regularization \cite{kirsten}, or radial WKB \cite{dunnemin}. The
finite part of these subtractions is related to the specific
renormalization prescription
\cite{strumia,baacke,dunnemin,dhlm,burnier}. In  \cite{dunnemin},
in flat space, it was checked explicitly that the result of this
procedure connects smoothly to the analytic thin-wall limit
results for the {\it renormalized}\/ fluctuation determinant.

A key element of this approach is knowledge of the large $l$
behavior of $\ln \left(\Det\left[ \Lambda^{(l)}\right]/{\Det\left[
\Lambda^{(l)}_{\rm free}\right]}\right)$, which must be subtracted
to make the $l$ sum finite (renormalization involves a further
step). In flat space the radial WKB analysis leads to the
following expression for the large $l$ behavior \cite{dunnemin}:
\begin{eqnarray}
\ln \left(\frac{\Det\left[ \Lambda^{(l)}\right]}{\Det\left[
\Lambda^{(l)}_{\rm free}\right]}\right) \sim
\frac{\frac{1}{2}\int_0^\infty dr \, r\,
W}{(l+1)}-\frac{\frac{1}{8}\int_0^\infty dr\, r^3\,
W(W+2V^{\prime\prime}[\phi_{\rm
fv}])}{(l+1)^3}+O\left(\frac{1}{(l+1)^5}\right) \quad, \qquad
l\to\infty\quad.
\label{wkbnlo}
\end{eqnarray}
Here
$W=V^{\prime\prime}[\phi]-V^{\prime\prime}[\phi_{\rm fv}]$.
Subtracting these terms makes the sum over $l$ in \eqn{formal}
finite, and is one part of the analysis leading to a finite
renormalized fluctuation determinant. We now turn our attention to
the large $l$ behavior of these determinants in the gravitational
case.

It is immediately clear that with gravitational coupling such a
computation cannot be done for the {\it renormalized} fluctuation
determinant, as we do not know how to renormalize gravity.
Nevertheless, we can study this question in the de Sitter limit,
where  the gravitational background is fixed to be of the de
Sitter form in \eqn{simple}. This limit is physically appropriate
when the variation of the potential on the scale of the barrier is
much less than $V_{\rm top}$ \cite{rubakov}. For the moderate
values of the cubic coupling $b$ in \eqn{potential} considered
here, this amounts to the condition $\beta\epsilon^2 \ll1$, which
means that the potential is large and positive, with:
\begin{equation}
H_{\rm fv}\approx H_{\rm top}\approx H_{\rm tv} \approx H\quad.
\label{H}
\end{equation}
Thus, fixing the metric to have the de Sitter form
\begin{equation}
a(\sigma)= \frac{1}{H}\sin\left(H\sigma\right)\quad,
\label{ds}
\end{equation}
the classical equations of motion reduce to a single equation for $\phi(\sigma)$:
\begin{equation}
\ddot{\phi}+3\frac{\dot{a}}{a}\dot{\phi}-V'(\phi)=0 \quad.
\label{dsphi}
\end{equation}
Even though $a(\sigma)$ is determined, one still finds various
different types of  oscillating bounce solutions for
$\phi(\sigma)$, as described in the previous section. These have
been extensively studied recently in \cite{hw} with a different
choice of parameter $\beta=70.03$.

To compute the determinant ratio in \eqn{gy}, consider first the
false vacuum case, which is the appropriate ``free'' reference
operator: $ \Lambda^{(l)}_{\rm free}= \Lambda^{(l)}_{\rm fv}$.
Since $\phi=\phi_{\rm fv}$ is constant, and $V^\prime(\phi_{\rm
fv})$ vanishes, the fluctuation potential \eqn{flucpot} simplifies
dramatically, and the Jacobi equation \eqn{sturm} for zero
eigenvalue becomes
\begin{equation}
-\ddot{\Phi}_{\rm fv}^{(l)}-3\frac{\dot{a}}{a}\dot{\Phi}_{\rm
fv}^{(l)}+\left[V''(\phi_{\rm
fv})+\frac{l(l+2)}{a^2}\right]\Phi_{\rm fv}^{(l)}=0 \quad,
\end{equation}
with $a(\sigma)=\frac{1}{H_{\rm fv}}\sin(H_{\rm fv} \sigma)$. The
zero mode solution with the correct initial value behavior,
$\Phi_{\rm fv}^{(l)}(\sigma)\sim \sigma^l$, is an associated
Legendre function (essentially derivatives of a conical function)
\begin{equation}
\Phi_{\rm fv}^{(l)}(\sigma)=\frac{N_{\rm fv}}{\sin\left(H_{\rm fv}\sigma\right)} P^{l+1}_{-\frac{1}{2}+i\sqrt{\frac{ V''(\phi_{\rm fv})}{H_{\rm fv}^2}-\frac{9}{4}}}\left[\cos\left(H_{\rm fv}\sigma\right)\right]\quad,
\label{conical}
\end{equation}
where $N_{\rm fv}$ is an unimportant normalization constant. This
function is positive definite and diverges as $H_{\rm fv}\,\sigma
\to \pi$.

Now consider the same computation but for a bounce solution.
Immediately we find a significant difference between the flat and
gravitational cases. In flat space both free and bounce solutions
are defined on the same interval $r\in [0, \infty)$. But with
gravity, a nontrivial bounce solution is defined on the interval
$[0, \sigma_{\rm max}^{\rm bounce}]$, where the interval is
determined by the second zero of the metric function $a(\sigma)$.
So, in general, for solutions of the full bounce equations
\eqn{phieq}-\eqn{bcs}, the false vacuum solution and a nontrivial
bounce are defined on different intervals. Fortunately, this
problem goes away precisely in the de Sitter limit being
considered here, where we can take the metric field to be of the
form in \eqn{ds} with $H=H_{\rm fv}$, so that both solutions live
on the same interval.

As in the flat space case \cite{baacke,dunnemin}, given that the
free solution \eqn{conical} is known analytically, it is better to
consider the  {\it ratio} of the functions appearing in \eqn{gy}:
\begin{eqnarray}
T^{(l)}(\sigma)\equiv
\frac{\Phi^{(l)}(\sigma)}{\Phi^{(l)}_{\rm fv}(\sigma)}\quad ,
\label{ratio}
\end{eqnarray}
because this ratio remains finite as $\sigma\to
\sigma_{\rm max}$. This ratio satisfies the following differential
equation with  simple initial value boundary conditions:
\begin{eqnarray}
-\ddot{T}-\left[2\frac{\dot{\Phi}_{\rm fv}}{\Phi_{\rm fv}} +
3\frac{\dot{a}}{a}-\frac{\dot{{\mathcal Q}}}{{\mathcal Q}}\right]\dot{T} + \left[{\mathcal Q}
U+\frac{\dot{{\mathcal Q}}}{{\mathcal Q}}\,\frac{\dot{\Phi}_{\rm fv}}{\Phi_{\rm fv}}-
U_{\rm fv}\right]T &=& 0 \quad,\nn T(0)=1\quad , \quad
\dot{T}(0)=0&&\quad\quad. \label{ratiogy}
\end{eqnarray}
This also shows why
the normalization of the false vacuum solution $\Phi_{\rm fv}$ in
\eqn{conical} is not important. It is conventional to compute the
logarithm of the determinant ratio, in which case the
Gel'fand-Yaglom result \eqn{gy} can be written simply as
\begin{eqnarray}
\ln\left(\frac{\Det\left[ \Lambda^{(l)}\right]}{\Det\left[
\Lambda^{(l)}_{\rm free}\right]}\right) =
\ln\left[T^{(l)}(\sigma_{\rm max})\right]\quad.
\label{gyr}
\end{eqnarray}

\begin{figure}[tb]
\begin{center}
\includegraphics[width=0.7\textwidth]{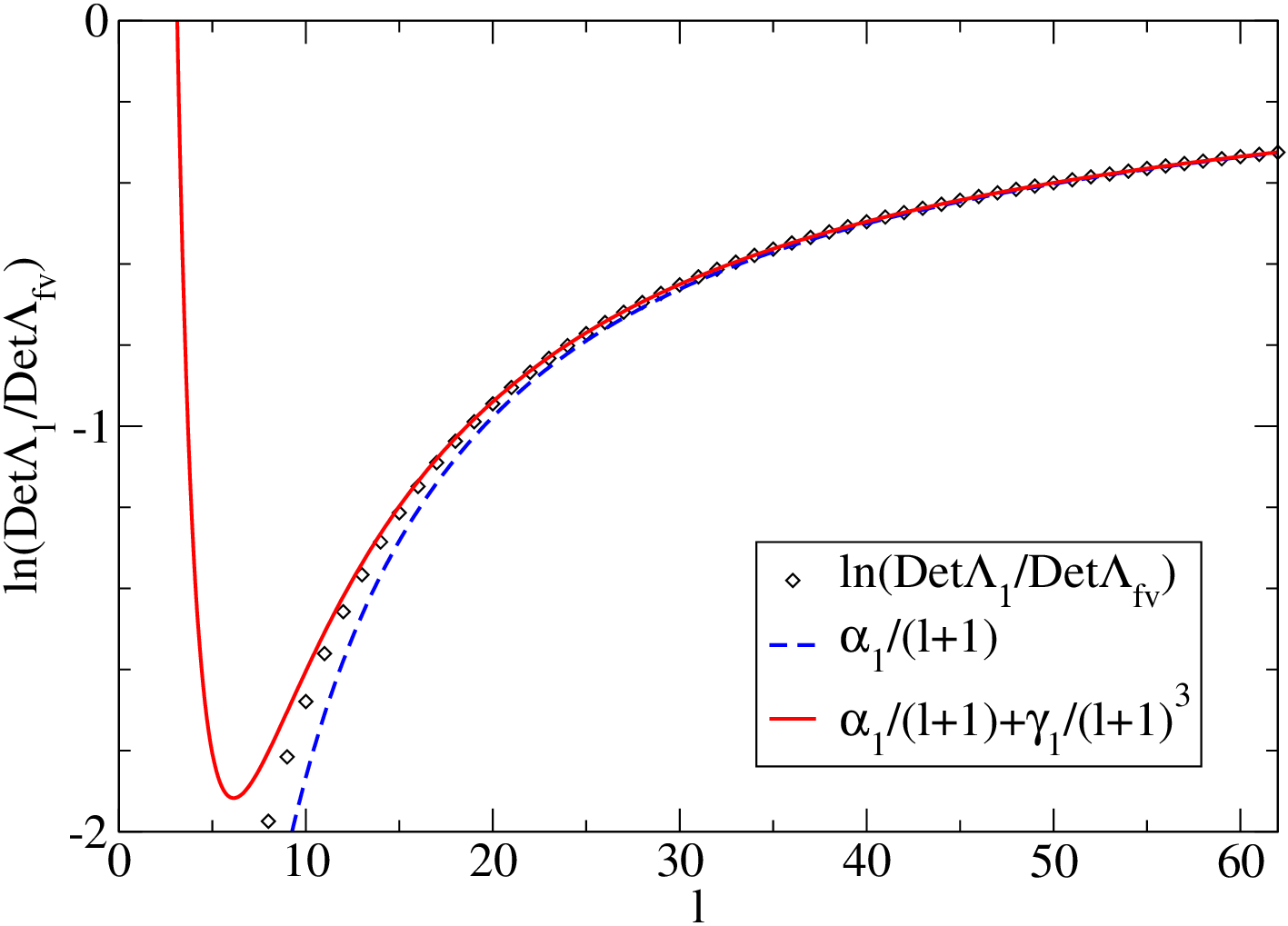}
\end{center}
\caption{$\ln\left[\Det \Lambda^{(l)}/\Det\Lambda^{(l)}_{\rm
free}\right]$ for the single bounce, for parameter values:
$\beta=45$, $b=0.25$, $\epsilon=0.046$, $\varphi(0)=\varphi_{\rm
tv}-0.000\,551\,736\,739\,278\,415\,5$, $\varphi(\pi)=\varphi_{\rm
fv}+0.000\,001\,929\,113\,943\,772\,119\,407\,698\,5$. Diamond
points denote the numerical results using the Gel'fand-Yaglom
technique as in \eqn{ratiogy}-\eqn{gyr}; the (blue) dash line
denotes the leading large $l$ behavior in \eqn{largel} with
$\alpha$ defined in \eqn{alpha} and $\gamma=0$; the (red) solid
line denotes the leading and subleading large $l$ behavior in
\eqn{largel} with $\alpha$ in \eqn{alpha} and $\gamma$ defined in
\eqn{gamma}.} \label{fig:dSl1}
\end{figure}

\begin{figure}[htb]
\begin{center}
\includegraphics[width=0.7\textwidth]{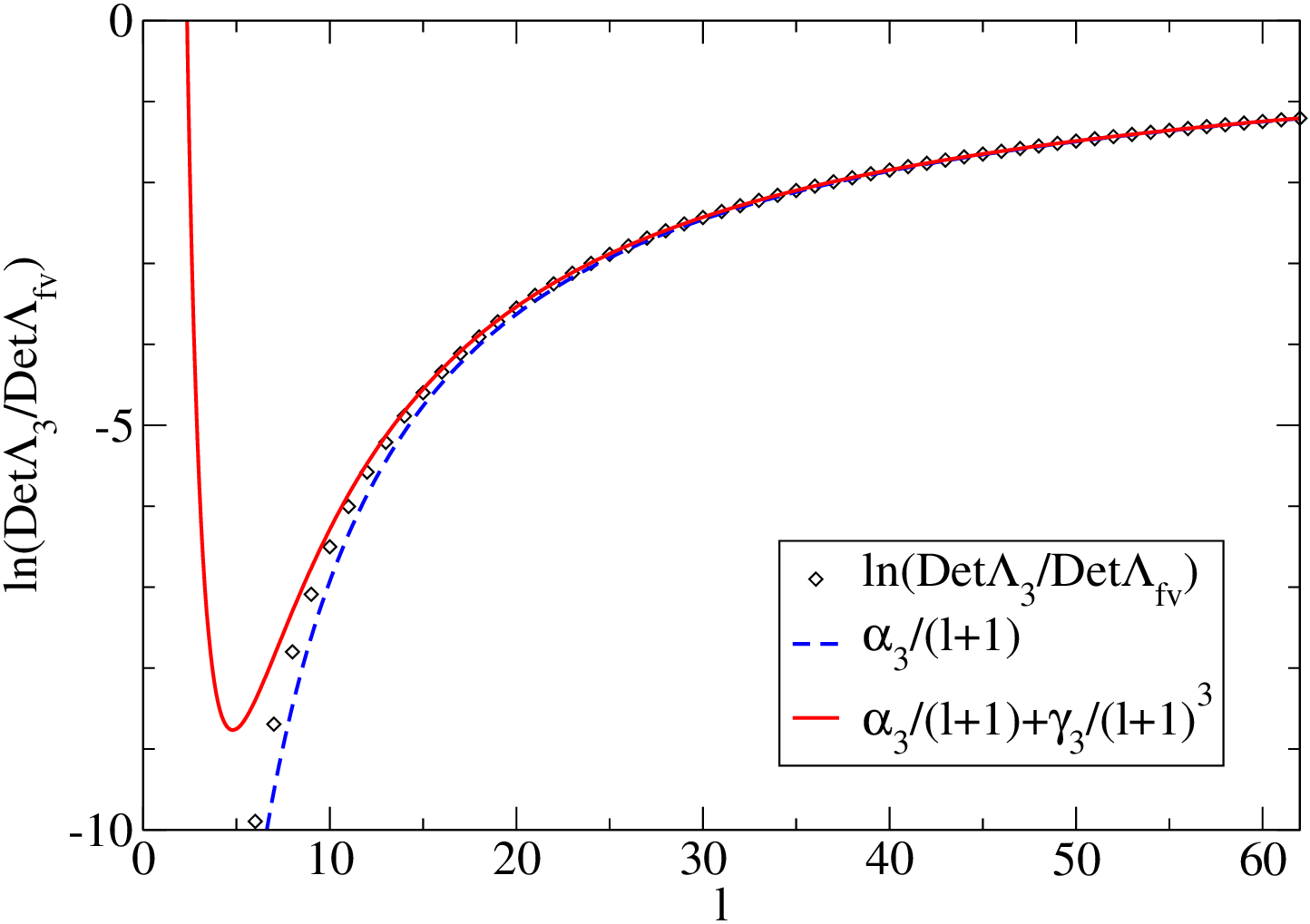}
\end{center}
\caption{$\ln\left[\Det \Lambda^{(l)}/\Det\Lambda^{(l)}_{\rm
free}\right]$ for the triple bounce, for parameter values:
$\beta=45$, $b=0.25$, $\epsilon=0.046$, $\varphi(0)=\varphi_{\rm
tv}-0.007\,517\,937\,804\,678\,529$, $\varphi(\pi)=\varphi_{\rm
fv}+0.001\,906\,935\,294\,979\,618$. Diamond points denote the
numerical results using the Gel'fand-Yaglom technique as in
\eqn{ratiogy}-\eqn{gyr}; the (blue) dash line denotes the leading
large $l$ behavior in \eqn{largel} with $\alpha$ defined in
\eqn{alpha} and $\gamma=0$; the (red) solid line denotes the
leading and subleading large $l$ behavior in \eqn{largel} with
$\alpha$ in \eqn{alpha} and $\gamma$ defined in \eqn{gamma}.}
\label{fig:dSl3}
\end{figure}

\begin{figure}[htb]
\begin{center}
\includegraphics[width=0.7\textwidth]{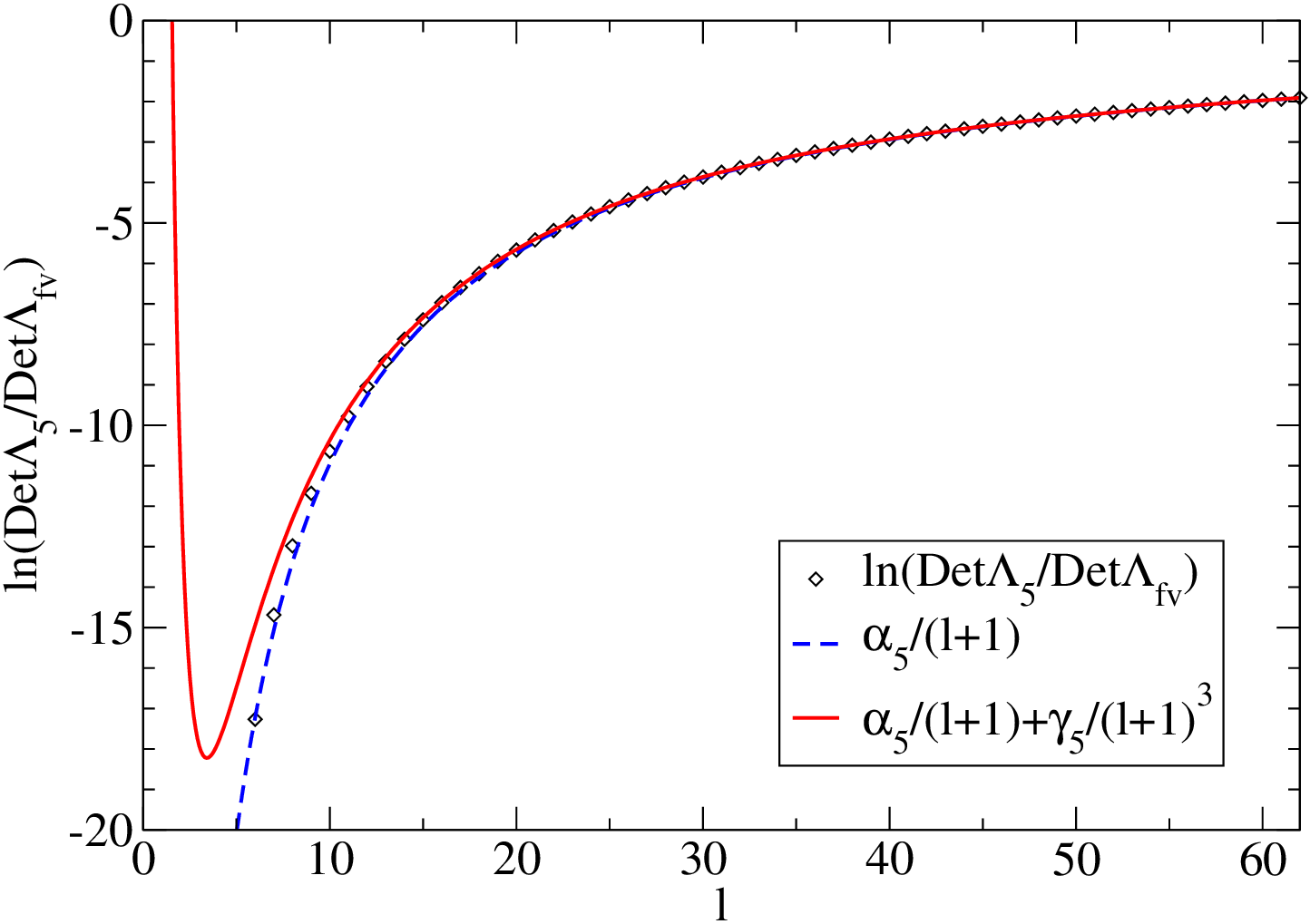}
\end{center}
\caption{$\ln\left[\Det \Lambda^{(l)}/\Det\Lambda^{(l)}_{\rm
free}\right]$ for the quintuple bounce, for parameter values:
$\beta=45$, $b=0.25$, $\epsilon=0.046$, $\varphi(0)=\varphi_{\rm
tv}-0.303\,787\,546\,677\,413\,1$, $\varphi(\pi)=\varphi_{\rm
fv}+0.187\,525\,348\,746\,153\,5$. Diamond points denote the
numerical results using the Gel'fand-Yaglom technique as in
\eqn{ratiogy}-\eqn{gyr}; the (blue) dash line denotes the leading
large $l$ behavior in \eqn{largel} with $\alpha$ defined in
\eqn{alpha} and $\gamma=0$; the (red) solid line denotes the
leading and subleading large $l$ behavior in \eqn{largel} with
$\alpha$ in \eqn{alpha} and $\gamma$ defined in \eqn{gamma}.}
\label{fig:dSl5}
\end{figure}

\begin{figure}[htb]
\begin{center}
\includegraphics[width=0.7\textwidth]{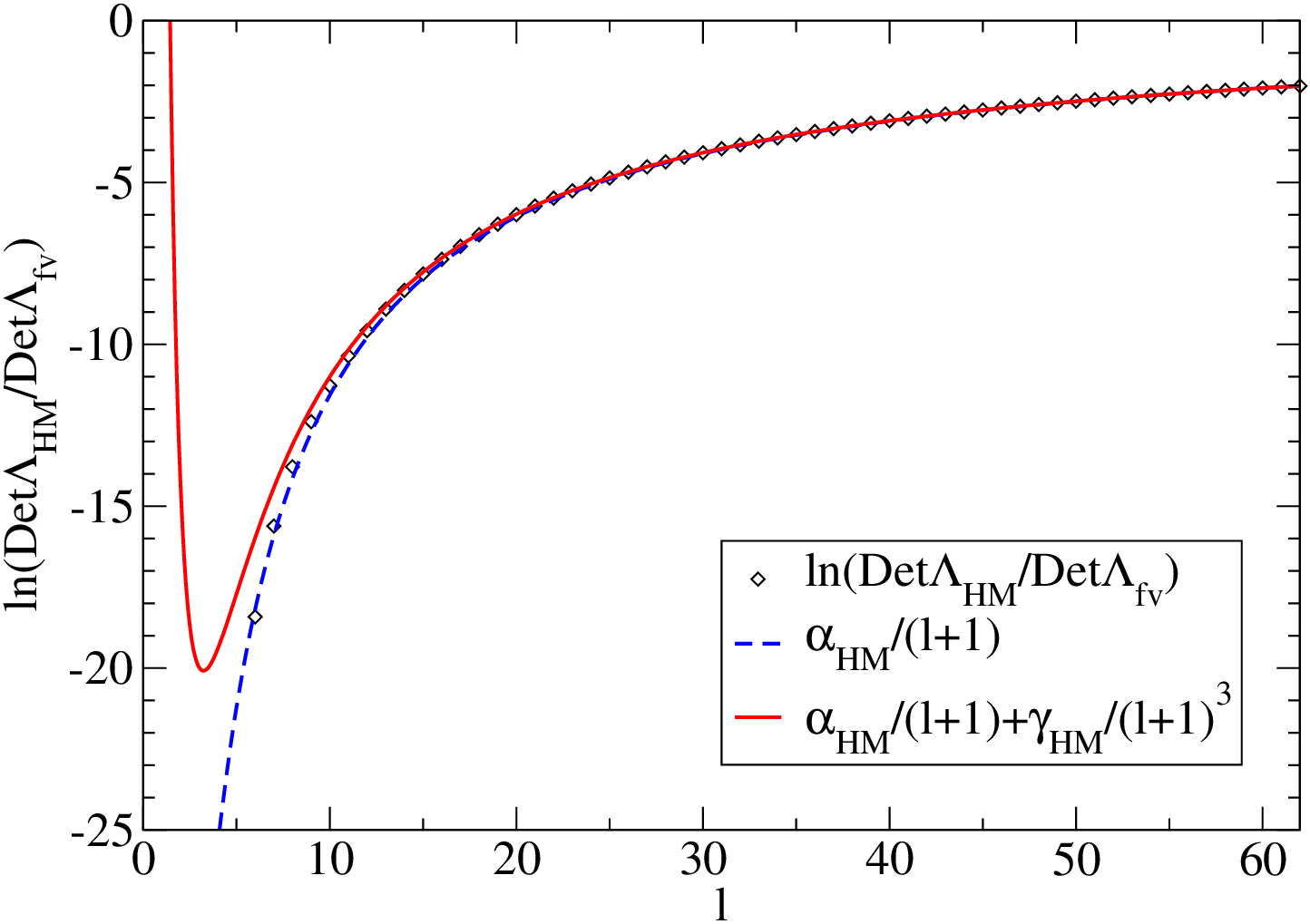}
\end{center}
\caption{$\ln\left[\Det \Lambda^{(l)}/\Det\Lambda^{(l)}_{\rm
free}\right]$ for the Hawking-Moss bounce, for parameter values:
$\beta=45$, $b=0.25$, $\epsilon=0.046$. Diamond points denote the
numerical result using the Gel'fand-Yaglom technique as in
\eqn{ratiogy}-\eqn{gyr}; the (blue) dash line denotes the leading
large $l$ behavior in \eqn{largel} with $\alpha$ defined in
\eqn{alpha} and $\gamma=0$; the (red) solid line denotes the
leading and subleading large $l$ behavior in \eqn{largel} with
$\alpha$ in \eqn{alpha} and $\gamma$ defined in \eqn{gamma}.}
\label{fig:dSlHM}
\end{figure}

We have computed these determinants as a function of $l$, by
numerically integrating the initial value problem \eqn{ratiogy}
for various bounce solutions, using the fluctuation equation in
\eqn{sturm}, with metric field given by the de Sitter form
\eqn{ds}, and the scalar field given by solutions to the bounce
equation \eqn{dsphi}. The scalar field bounce solutions in this de
Sitter case with same parameter $\beta=45$ are very close to those
shown in Figures \ref{fig:phi1}, \ref{fig:phi3} and
\ref{fig:phi5}, so we do not bother plotting them again.  The
results for the logarithm of the determinant ratio are shown as
diamond points in Figures \ref{fig:dSl1} through \ref{fig:dSlHM}.
The first empirical observation is that the large $l$ behavior is
very similar to that \eqn{wkbnlo} found in the flat space case
\cite{dunnemin}:
\begin{equation}
\ln\left[\frac{\Det(\Lambda^{(l)})}{\Det(\Lambda_{\rm
free}^{(l)})}\right]
\sim\frac{\alpha}{l+1}+\frac{\gamma}{(l+1)^3}+\dots \quad, \qquad
l\to\infty\quad, \label{largel}
\end{equation}
To extract the leading coefficient $\alpha$, we consider the
leading and subleading large $l$ behavior of the Jacobi equation
\eqn{sturm}:
\begin{equation}
-\ddot{\Phi}^{(l)} -
3\frac{\dot{a}}{a}\dot{\Phi}^{(l)}
+ \left[V''(\phi) + 2\kappa\dot{\phi}^2 +
\frac{l(l+2)}{a^2}\right]\Phi^{(l)} = \lambda^{(l)}
\Phi^{(l)}\quad, \qquad (l\gg1)\quad. \label{leadingjacobi}
\end{equation}
Adapting the WKB analysis of \cite{dhlm,dunnemin} leads to the
following result for the leading large $l$ dependence of the log
of the determinant ratio:
\begin{eqnarray}
\alpha= \frac{1}{2}\int_0^{\sigma_{\rm max}}d\sigma\,
a(\sigma)\left[V''(\phi)+2\kappa\dot{\phi}^2- V''(\phi_{\rm
fv})\right]\quad, \label{alpha}
\end{eqnarray}
Notice the close similarity to the leading term in the flat space
large $l$ behavior in \eqn{wkbnlo}. This curved-space leading
behavior $\alpha/(l+1)$, with $\alpha$ given by \eqn{alpha}, is
shown in Figures \ref{fig:dSl1} through \ref{fig:dSlHM} as dashed
(blue) curves, and we see that the agreement at large $l$ is very
good. It is much harder to find the next-to-leading behavior
because the subleading $l$ dependence of the Jacobi equation is
very complicated. Nevertheless, by analogy with the flat space
case \eqn{wkbnlo} we propose the estimate
\begin{eqnarray}
\gamma\approx -\frac{1}{8}\int_0^{\sigma_{\rm max}}d\sigma\,
a^3(\sigma)\left[V''(\phi)+2\kappa\dot{\phi}^2 - V''(\phi_{\rm
fv})\right]\left[V''(\phi)+2\kappa\dot{\phi}^2 + V''(\phi_{\rm
fv})\right]. \label{gamma}
\end{eqnarray}
Including this subleading behavior in \eqn{largel} produces the
solid (red) curves in Figures \ref{fig:dSl1} through
\ref{fig:dSlHM}, and we see that the agreement with the exact
numerical results is noticeably improved and is excellent at large
$l$, and is characteristic of an asymptotic large $l$ expansion in
its behavior at small $l$. Thus, the estimate in \eqn{gamma} is
very close to the exact answer. We found similar behavior for
other bounces.

\section{Negative modes}
\label{neg-sec}

In this section we turn to another important property of the
fluctuation operator, namely the existence of negative modes.
Here, to be more general we can return to the general bounce
solutions, not needing to work in the de Sitter limit any more
(although we find the results to be the same in either case). In
the flat space false vacuum decay problem, it has been shown that
the fluctuation problem \eqn{flatsturm} has one and only one
negative mode, and that this occurs in the $l=0$ sector
\cite{coleman3}. This single negative mode plays an important
physical role in the semiclassical quantization, accounting for
the decaying nature of the process \cite{langer,stone,coleman1}.
In the gravitational case, there has been considerable effort
analyzing the appearance of negative modes in the $l=0$
sector \cite{tanaka,turok1,turok2,lav1,lav2,lav3}. For the scalar
field fluctuations characterized by the secondary action
\eqn{second}, the oscillating $n$-bounce solution has $n$ negative
modes in the $l=0$ sector \cite{lav3}. Here we show that the
higher $n$ oscillating bounce solutions also can have negative
modes in higher $l$ sectors, while the single bounce solution, the
Coleman-De Luccia solution, has precisely one negative mode, which
is for $l=0$.

A direct numerical method for counting negative modes is based on
an important theorem in the calculus of variations due to Morse
\cite{morse}. It states that the number of negative modes of the
fluctuation operator $\Lambda$ is given by the number of zeros of
the solution of the initial value Jacobi equation $\Lambda\Phi=0$.
This is  consistent with the related Gel'fand-Yaglom result
\eqn{gy} for the computing the determinant as the value of
$\Phi(\sigma_{\rm max})$, since an odd number of zeros leads to a
negative determinant. This Morse analysis has been applied to the
counting of negative modes in the flat space false vacuum decay
problem in \cite{maziashvili}.

\begin{table}[htb]
\caption{The number of negative modes without including the
degeneracy factor $(l+1)^2$, for the parameter values: $\beta=45$,
$\epsilon=0.23$.}
\begin{tabular}{cccccccc}
\hline
\hline
$l$ &~0~&~2~&~3~&~4~&~5~&~6~&~7\\
\hline
1-bounce &1&0&0&0&0&0&0\\
\hline
3-bounce &3&2&2&1&1&0&0\\
\hline
5-bounce &5&4&3&2&1&0&0\\
\hline
HM &6&4&3&2&1&0&0\\
\hline
\hline
\end{tabular}
\label{tab1}
\end{table}

So to count the number of negative modes for a given bounce
solution, we numerically integrate the fluctuation Jacobi equation
\eqn{sturm}, with initial value boundary condition
$\Phi^{(l)}\sim\sigma^l$, and count how many times this function
changes sign on the interval $[0, \sigma_{\rm max}]$. In this case
we can do this computation using the full bounce solutions as
obtained in Section \ref{bounce-sec}, not just in the de Sitter
limit where the metric function $a(\sigma)$ is chosen to take the
sinusoidal form \eqn{ds}. In this way, we confirm the results of
\cite{turok2,lav3} that the $n$-bounce solution has $n$ negative
modes in the $l=0$ sector. More surprisingly, we find that for
$l\geq 2$ there are some negative modes for the higher $n$
oscillating bounce solutions. The precise pattern depends on the
parameters in the potential, but a representative counting is
shown in Table \ref{tab1}. As the oscillation number of the bounce
increases there are more negative modes, and they extend to higher
values of $l$. In studying many single-bounce solutions we have always 
found only one negative mode, and always in the $l=0$ sector. We
also found the same negative mode counting pattern using
Lavrelashvili {\it et al}\/'s fluctuation operator in
\eqn{lavpot}.

To put this result of extra negative modes at higher $l$ in some
perspective, consider the Hawking-Moss solution, which is the
large $n$ limit of the $n$-bounce solution \cite{hw}. Here, we can
write the exact solution to the zero eigenvalue Jacobi equation,
$\Lambda \Phi=0$, with initial behavior $\Phi\sim\sigma^l$, as an
associated Legendre function, analogous to the false vacuum
solution in \eqn{conical}:
\begin{equation}
\Phi_{\rm HM}^{(l)}(\sigma)=\frac{N_{\rm HM}}{\sin\left(H_{\rm
top}\sigma\right)}
P^{l+1}_{-\frac{1}{2}+\sqrt{\beta+\frac{9}{4}}}\left[\cos\left(H_{\rm
top}\sigma\right)\right]\quad.
\end{equation}
Here $\beta$ is the parameter defined in \eqn{beta}, and $N_{\rm HM}$ is an unimportant normalization constant. The
counting of the zeros of this function can be done precisely, and
one finds that the number of zeros depends critically on the value
of $\beta$. For $N(N+3)<\beta\le(N+1)(N+4)$, there are $N+1$
zeros. By the Morse theorem the number of nodes of this zero mode
solution is equal to the number of negative modes in the
perturbation. This counting is also shown as the last row in Table
\ref{tab1}, and we have also confirmed that the numerical
integration and the exact result give the same counting. Since the
oscillating bounce solutions tend to this Hawking-Moss solution,
this goes some way towards explaining the origin of these new
negative modes for the oscillating bounce solutions at higher $l$.
Physically, this is extra evidence that these higher $n$
oscillating bounce solutions are not directly related to quantum
tunneling, as suggested already in \cite{hw}.

\section{Conclusion}
\label{conclusions-sec}

In this paper we have analyzed several issues concerning the
quantum fluctuations about classical bounce solutions in the
theory of a self-interacting scalar field interacting with
gravity. In flat space the semiclassical fluctuation analysis can
be done completely, both analytically in the thin-wall limit and
numerically for more general potentials. In the gravitational
case, the fluctuation problem still separates into a set of
one-dimensional fluctuation problems labeled by an integer $l$. We
found the leading large $l$ behavior \eqn{largel} - \eqn{alpha},
and an estimate \eqn{gamma} for the subleading behavior, of the
logarithm of the determinant of the fluctuation operator. The
agreement with the numerical computations is impressive. We also
analyzed the existence of negative modes using Morse's theorem,
confirming that the single-bounce Coleman-De Luccia solution has a
single negative mode, which lies in the $l=0$ sector, and that the
oscillating $n$-bounce solution has $n$ negative modes in the
$l=0$ sector. We also found new negative modes for the oscillating
$n$-bounce solutions for higher $n$ with $l\geq 2$. This adds
further weight to the physical interpretation suggested in
\cite{hw} that these bounces are not directly related to quantum
tunneling, but rather are related to the thermal character of
quantum field theory in de Sitter space, and interpolate to the
Hawking-Moss solution for large $n$.

Many problems remain. The standard scalar fluctuation analysis
\cite{turok1,turok2,lav1,lav2} in the gravitational case precludes
consideration of the $l=1$ sector, and so we cannot yet say
anything about the collective coordinate contribution to the
renormalized fluctuation determinant. In the flat space case it
was recently shown how this $l=1$ contribution, combining the
determinant with the zero modes removed and the collective
coordinate contribution, could be expressed simply in terms of the
asymptotic properties of the classical bounce solution
\cite{dunnemin}. Whether something like this can be found for the
gravitational case depends on a different analysis of the $l=1$
fluctuation problem. Perhaps the most challenging problem is that
the renormalization of quantum gravity is not understood. In the
flat space case, without gravity, the subtractions made from the
regularized determinant for each $l$ include a finite piece
depending on the regularization scale. These can be associated
with renormalization, permitting the computation of a {\it finite
and renormalized} fluctuation determinant \cite{baacke,dunnemin}.
An important preliminary step for the gravitational case would be
to develop fully this approach in the limit where the
gravitational background is fixed to be de Sitter, in which case
the large $l$ behavior of the log determinants is given by
\eqn{largel}, and where the perturbative renormalization of the
scalar field in a fixed curved background is known \cite{birrell}.
Hopefully this can shed further light on the important question of
the nature of the semiclassical path integral approximation in the
presence of de Sitter gravity \cite{rubakov,garriga,vachaspati}.

\vskip .5cm
{\bf Acknowledgments:}  We thank the US DOE for
support through the grant DE-FG02-92ER40716.

\appendix\section{Related forms of the fluctuation operator}
\label{appendix-sec}

In discussing the existence of negative modes it is
possible to make other choices than \eqn{weight} for the weight
function, yielding superficially  different-looking Jacobi
equations \cite{turok1,turok2,lav1,lav2,lav3}. But for the
purposes of computing the determinant, where the magnitude of the
eigenvalues is also relevant, the choice in \eqn{sturm} is the
most direct. For completeness, the choice of Lavrelashvili {\it et
al}\/ is to use the perturbation function
$\Phi_L\equiv\sqrt{a^3/{\mathcal Q}}\delta\phi$, and weight function $a^3/{\mathcal Q}$,
leading to the Jacobi equation \cite{lav1,lav2,lav3}
\begin{equation}
-\frac{d^2\Phi_L}{d\sigma^2}+U_L[a,\phi]\Phi_L=\lambda_L \Phi_L\quad,
\label{lavsturm}
\end{equation}
where
\begin{eqnarray}
U_L[a,\phi]  &\equiv& \frac{1}{{\mathcal Q}}V''(\phi) - \frac{3\kappa
a^2}{2{\mathcal Q}^2(-\Delta_3-3K)}\left[V'(\phi)\right]^2 + \frac{6 \kappa
a\dot{a}\dot{\phi}}{{\mathcal Q}^2(-\Delta_3-3K)}V'(\phi) -
\frac{\kappa}{6}\left[\dot{\phi}^2+V(\phi)\right] \nn &&+
\frac{\dot{a}^2}{a^2}\left(-\frac{1}{4}-\frac{10}{{\mathcal Q}} +
\frac{12}{{\mathcal Q}^2}\right) + \frac{1}{a^2}\left[(-\Delta_3-3K)({\mathcal Q}+2) -
\frac{-2\Delta_3-8K}{{\mathcal Q}}\right]\quad,
\label{lavpot}
\end{eqnarray}
which agrees with Eqn.~(19) in the second reference in \cite{lav3} when $K=1$ and $\Delta_3=0$.
Furthermore, this potential satisfies
\begin{equation}
U_L[a,\phi]= {\mathcal Q}\, U[a,\phi] +
\sqrt{\frac{{\mathcal Q}}{a^3}}\frac{d^2}{d\sigma^2}\sqrt{\frac{a^3}{{\mathcal Q}}}\quad.
\label{lavcomparison}
\end{equation}
Using this fluctuation operator, we found the same pattern of negative modes
shown in Table \ref{tab1}.

Another choice, yielding an elegant result for the existence of
negative modes, was made by Turok {\it et al}, who chose the
perturbation function  $\Phi_T\equiv\delta\phi/\dot{\phi}$, and
weight function $1/\dot{\phi}^2$, leading to the Jacobi equation
\cite{turok1,turok2}
\begin{equation}
-\frac{d}{d\sigma}\left(\frac{a^3\dot{\phi}^2}{{\mathcal Q}}\frac{d
\Phi_T}{d\sigma}\right) + U_T[a,\phi]\Phi_T=\lambda_T \Phi_T\quad,
\label{turoksturm}
\end{equation}
Here
\begin{equation}
U_T[a,\phi]\equiv a\, \dot{\phi}^2(-\Delta_3-3K)\quad,
\label{turokpot}
\end{equation}
and it is related to our $U[a, \phi]$ by
\begin{equation}
U_T[a,\phi]= a^3\dot{\phi}^2U[a,\phi] - \dot{\phi}
\frac{d}{d\sigma}\left(\frac{a^2\ddot{\phi}}{{\mathcal Q}}\right)\quad.
\label{turokcomparison}
\end{equation}

\end{document}